\documentclass[10pt,amsmath,amssymb,pra,showpacs,twocolumn]{revtex4-1}
\usepackage{mathtools}
\usepackage{enumerate}
\usepackage{graphicx}
\usepackage{dcolumn}
\usepackage{bm}
\usepackage{braket}
\usepackage{booktabs}
\usepackage{color}
\newcounter{one}
\def\ketbra#1#2{\ket{#1}\!\bra{#2}}

\newcommand{\Tr}[0]{ \mathrm{Tr}}

\newtheorem{theorem}{Theorem}

\newtheorem{definition}{Definition}

\def\QED{\mbox{\rule[0pt]{1.5ex}{1.5ex}}}

\def\endproof{\hspace*{\fill}~\QED\par\endtrivlist\unskip}

\newcommand\calE{{\cal E}}

\newcommand{\beq}{\begin{equation}}
\newcommand{\eeq}{\end{equation}}

\begin{document}
\title{Fisher information matrix as a resource measure in resource theory of asymmetry with general connected Lie group symmetry}
\author{Daigo Kudo}
	\email{daigo09015@gmail.com}
\author{Hiroyasu Tajima$^{1,2}$}
	\email{hiroyasu.tajima@uec.ac.jp}
\affiliation{
				1. Graduate School of Informatics and Engineering, The University of Electro-Communications,1-5-1 Chofugaoka, Chofu, Tokyo 182-8585, Japan
			}
\affiliation{
2. JST, PRESTO, 4-1-8 Honcho, Kawaguchi, Saitama, 332-0012, Japan
			}

\begin{abstract}
In recent years, in quantum information theory, there has been a remarkable development in the general theoretical framework for studying symmetry in dynamics.
This development, called resource theory of asymmetry, is expected to have a wide range of applications, from accurate time measurements to black hole physics.
Despite its importance, the foundation of resource theory of asymmetry still has room for expansion. An important problem is in quantifying the amount of resource.
When the symmetry prescribed U(1), i.e., with a single conserved quantity,  the quantum Fisher information is known as a resource measure that has suitable properties and a clear physical meaning related to quantum fluctuation of the conserved quantity.
However, it is not clear what is the resource measure with such suitable properties when a general symmetry prevails for which there are multiple conserved quantities.
The purpose of this paper is to fill this gap.
Specifically, we show that the quantum Fisher information matrix is a resource measure whenever a connected linear Lie group describes the symmetry.
We also consider the physical meaning of this matrix and see which properties that the quantum Fisher information has when the symmetry is described by $U(1)$ can be inherited by the quantum Fisher information matrix.  
\end{abstract}

\maketitle

\section{Introduction}
Resource theory measures quantum resources (e.g., entanglement and coherence) and considers the difference between what tasks are possible with and without the resources. The generality of resource theory enables us to analyze fundamental physical problems in resource-theoretical settings.
Within this background, resource theory has produced many variants \cite{QT1,QT2,RToC0,RToC1,Marvian_thesis,skew_resource,Takagi_skew,Marvian_distillation,YT,WAY_RToA1,WAY_RToA2,TN,TSS,TSS2,TS,Marvian2018,Lostaglio2018,TSL_RToA,e-EKZhou,e-EKYang,Liu1,Liu2}.
These variants differ in the types of situations and resources they treat, depending on the target problem, and each of them constitutes a field in quantum physics.
One of the most fundamental variants is the resource theory of asymmetry (RTA) \cite{Marvian_thesis,skew_resource,Takagi_skew,Marvian_distillation,YT,WAY_RToA1,WAY_RToA2,TN,TSS,TSS2,TS,Marvian2018,Lostaglio2018,TSL_RToA,e-EKZhou,e-EKYang,Liu1,Liu2}, the subject of this paper.

RTA is a resource theory that treats symmetries of dynamics  \cite{Marvian_thesis,skew_resource,Takagi_skew,Marvian_distillation}.
Because it is related to symmetry, a fundamental concept in physics, it has a wide range of applications including constructing accurate clocks \cite{Marvian_distillation,woods_autonomous_2019}, accurate measurements on non-conserved quantities \cite{WAY_RToA1,WAY_RToA2,TN}, unitary gate implementation \cite{TSS,TSS2,TS}, coherence distribution \cite{Marvian2018,Lostaglio2018}, quantum speed limit \cite{TSL_RToA}, quantum error correction \cite{TS,e-EKZhou,e-EKYang,Liu1,Liu2}, and black hole information loss problem \cite{TS}.
Despite these applications, the foundations of RTA still have several  parts that can be extended.
One such part is in quantifying resources.
In resource theories, we often try to clarify the limitations of performance given a certain amount of a resource.
Therefore, quantifying resources is important in every variant of resource theory. 
Given its importance, a quantification has also been done in RTA \cite{skew_resource,Takagi_skew,Marvian_distillation,YT}.
When the symmetry is described by U(1), i.e., with a single conserved quantity, a good resource measure is known. 
Quantum Fisher information \cite{Helstrom} satisfies the suitable properties of a resource measure and has a clear physical meaning related to quantum fluctuation of the conserved quantity corresponding to the symmetry \cite{skew_resource,Takagi_skew,Marvian_distillation}.
Quantum Fisher information is also useful in constructing a theory for the convertibility of pure states in the non-i.i.d. (independent and identically distributed) regime in RTA \cite{YT}.
However, it is not clear what the resource measure with similar properties is for general symmetries, for which there are multiple conserved quantities.

In this paper, we show that the quantum Fisher information matrix is a resource measure whenever the symmetry is described by a connected linear Lie group. 
Here, all the basic properties of the resource measure, including monotonicity, are given as matrix inequalities.
So far, resource measures have never been given as matrices, and our result is the first example in resource theory where a matrix provides a resource measure.
Our result also corresponds to the multivariable extension of the results that Fisher information is a resource measure if the symmetry described by U(1) \cite{skew_resource,Takagi_skew,Marvian_distillation}.
If we restrict the symmetry to that described by U(1), our results recover the known results produced so far.

We also consider the physical meanings of the quantum Fisher information matrix.
In each variant of resource theory, each resource has a physical meaning much like entanglement means a type of quantum correlation.
Quantum Fisher information, which is a resource measure when the symmetry is described by U(1), relates to the quantum fluctuation of the conserved quantity corresponding to the symmetry \cite{min_V_Petz,min_V_Yu,Luo,Hansen,Marvian_distillation}.
In this paper, we extend the above meaning of quantum Fisher information to the quantum Fisher information matrix and give three results.

The first result is a proof that the quantum Fisher information matrix satisfies the conditions that characterize quantum fluctuation.
There are several criteria given by Luo for the characterization of quantum fluctuations \cite{Luo}, and it is known that Fisher information satisfies them \cite{skew_resource,Takagi_skew,GII}. We extend Luo's criteria to matrices and show that the quantum Fisher information matrix satisfies these extended criteria.
The second result is a comprehension of the relationship between the quantum Fisher information of a mixed state and the covariance matrix of its purification.
It is known that the quantum Fisher information of a mixed state is equal to the minimum of the variance of its purification \cite{Marvian_distillation}.
We extend this result to situations having multiple conserved quantities. 
As a result, we show that the quantum Fisher information matrix is equal to four times the minimum of the covariance matrix of its purification.
The third result is the relationship between the quantum Fisher information matrix of the mixed state and the average of the covariance matrix when decomposing the mixed state into its pure state.
It is also known that quantum Fisher information is equal to four times the minimum average of the variances \cite{min_V_Petz,min_V_Yu}.
We show by giving a counterexample that this theorem does not hold in the case of multiple conserved quantities.

The structure of this paper is as follows.
In section \ref{preliminary}, we overview the elementary basics required for this paper.
In particular, we review that, given U(1) symmetry, the quantum Fisher information is a resource measure, and is a measure of the quantum fluctuation of the conserved quantity.
In section \ref{results}, we show the main results in this paper.
Specificaly, the quantum Fisher information matrix is a resource measure whenever symmetry is described by a connected linear Lie group and then consider how other properties of the quantum Fisher information are inherited by the quantum Fisher information matrix.
In section \ref{conclusion}, we summarize this paper.
All proofs of the theorems in section \ref{results} are given in the appendix.

\section{Preliminary}\label{preliminary}
In this section, we outline the prerequisites required in this paper.
\subsection{Resource theory}
In resource theory, we define ``free states" as states that we can easily prepare and ``free operations" as operations that we can easily perform. We consider ``resource states" as states that we cannot realize by combining free states and free operations.
Depending on the definition of the free state and free operation, resource theory treats different resources.
For example, suppose we define free states as separable states and free operations as local operations and classical communications (LOCC).
In this scenario, the resource is called entanglement, and quantum teleportation can be performed using it.

To consider what operations are possible with a certain amount of a resource, knowing how much of that resource is available in a state is essential.
We call a function a resource measure when the function of a quantum state $R(\rho)$ satisfies the following conditions \cite{resource-review}:
\begin{enumerate}
  \item \(R(\rho)\ge0\)
  \item \(R(\rho)=0\Longleftrightarrow\)\(\rho\) is free state
  \item \(\mathcal{E}\) is free operation \(\Longrightarrow R(\rho) \ge R(\mathcal{E}(\rho))\).
\end{enumerate}
The first condition means that the function $R$ assesses the amount of the resource.
The second condition implies that a free state is a state without resources, and the third condition implies that a free operation is a quantum operation that does not increase the resource.

\subsection{Groups and symmetry}
When a system's properties are invariant under a class of operations, e.g., rotations and translations, we say the system is symmetric under the operations. In general, symmetries of a system are described by a group, which we now define.
\begin{definition}
  A group is a set $G$ together with a binary operation on $G$ that combines any two elements $a,b\in G$ to form an element $ab\in G$ such that following three requirements are satisfied:  
  \begin{enumerate}
      \item \(\forall a,b,c\in G,\ (ab)c=a(bc)\)
      \item \(\exists e\in G,\forall a\in G,\ ae=ea=a\)
      \item \(\forall a\in G,\exists a^{-1}\in G,\ a^{-1}a=aa^{-1}=e\).
  \end{enumerate}
\end{definition}

When a quantum system is isolated from the environment, its time evolution is described by a unitary operator.
That is, a state $\rho$ on the system evolves as \(\rho\rightarrow U\rho U^{-1}\) where $U$ is a unitary operator that acts on the Hilbert space describing the system. 
When the isolated quantum system has symmetry described by a group $G$, the unitary dynamics $U$ is an element of the projective unitary representationof the group $G$.
A projective unitary representationof group $G$ on a Hilbert space \(\mathcal{H}\) is a map from group $G$ to the Hilbert space $\mathcal{H}$ satisfying 
\begin{equation}\label{eq:projective-representation}
    U(g_1)U(g_2)=e^{i\omega(g_1,g_2)}U(g_1 g_2).
\end{equation}
In particular, if \(e^{i\omega(g_1,g_2)}=1\) holds for any $g_1,g_2\in G$, then \(U(g)\) is a unitary representation.
In general, situations where the group $G$ is U(1) or $\mathbb{R}$ correspond to situations where there is one conserved quantity. In contrast, symmetries described by the more general group $G$ correspond to situations with multiple conserved quantities.

\subsection{Linear Lie groups and Lie rings}
When a system has continuous symmetry, a linear Lie group describes the symmetry.
Let \(M(n,\mathbb{C})\) be the set of complex matrices and \(GL(n,\mathbb{C})\coloneqq\{X\in M(n,\mathbb{C})\mid\det X \neq 0\}\) be the set of regular matrices, which is called a general Lie group. 

\begin{definition}
    A linear Lie group is defined as a closed subgroup of the general Lie group $GL(n,\mathbb{C})$
\end{definition}
A closed set implies that any sequences of points in \(G\) converge to the element in \(G\).
The unitary group of degree $n$, denoted $U(n)$, is a linear Lie group for which the elements satisfy $uu^\dagger=1=u^\dagger u$ for all $u\in U(n)$.

\begin{definition}
  A Lie ring $V$ is a vector space over real numbers $\mathbb{R}$ or complex numbers $\mathbb{C}$ together with a binary operation $[\cdot,\cdot]\colon V\times V\to V$ called the Lie bracket satisfying
  \begin{align}
      &[x+y,z]=[x,z]+[y,z]\\
      &[ax,y]=a[x,y],\quad \forall a\in \mathbb{R}\ (\mathrm{or} \ a\in\mathbb{C})\\
      &[x,y]=-[y,x]\\
      &[x,[y,z]]+[y,[z,x]]+[z,[x,y]]=0.
  \end{align}
  If the vector space $V$ is a real vector space, $V$ is called a real Lie ring.
\end{definition}
Defining a set $\mathfrak{g}$ as $\mathfrak{g}:=\{X\in M(n,\mathbb{C})\mid\forall t,\exp tX\in G\}$ for a Lie group $G$, the set $\mathfrak{g}$ is a real Lie ring and called the Lie ring of linear Lie group $G$ for which the Lie bracket $[\cdot,\cdot]$ is defined as $[X,Y]=YX-XY$.
For example, the Lie ring of a general linear Lie group $GL(n,\mathbb{C})
$ is \(\mathfrak{gl}(n,\mathbb{C})=M(n,\mathbb{C})\) and the Lie ring of the unitary group $U(n)$ is \(\mathfrak{u}(n)=\{X\in M(n,\mathbb{C})\mid X^\dagger=-X\}\).

A representation of a Lie ring $\mathfrak{g}$ is a homomorphism from the Lie ring $\mathfrak{g}$ to the Lie ring $\mathfrak{gl}(n,\mathbb{C})$, formally, a map \(\Phi:\mathfrak{g}\to\mathfrak{gl}(n,\mathbb{C})\) satisfying three conditions:
\begin{align}
    &\Phi(aX)=a\Phi(X)\\
    &\Phi(X+Y)=\Phi(X)+\Phi(Y)\\
    &\Phi([X,Y])=[\Phi(X),\Phi(Y)].
\end{align}
The following theorem implies that a representation of a linear Lie group \(G\) naturally yields a representation of its Lie ring \(\mathfrak{g}\).
\begin{theorem}\label{Th-differential-representation}
  Let $G$ and $H$ be linear Lie groups and \(\Phi:G\to H\) be a continuous homomorphism. And let $\mathfrak{g}$ and $\mathfrak{h}$ be the Lie rings of the linear Lie groups $G$ and $H$, respectively.
  Then, there exists a unique linear map \(L(\Phi)=\varphi\colon\mathfrak{g}\to \mathfrak{h}\) satisfying \(\exp\varphi(X)=\Phi(\exp X)\).
  Furthermore, the following holds:
  \begin{enumerate}
      \item \(\varphi([X,Y])=[\varphi(X),\varphi(Y)]\)
      \item \(\left.\varphi(X)=\frac{d}{dt}\Phi(\exp tX)\right|_{t=0}\).
  \end{enumerate}
\end{theorem}
Next, we consider a connected linear Lie group.
\begin{definition}
  A set $G$ is said to be path-connected if, for any $x$ and $b$ in $G$, there exists a continuous function \(f:[0,1]\to G\) such that \(f(0)=x\) and \(f(1)=y\).
\end{definition}
If a set $G$ is a linear Lie group, connectedness and path-connectedness are equivalent.
The following theorem holds for a connected linear Lie group \(G\).
\begin{theorem}\label{Th-connected}
  For a linear Lie group, the following conditions are equivalent:
  \begin{enumerate}[(i)]
      \item \(G\) is connected.
      \item \(G\) is generated by \(\exp \mathfrak{g}\).
      That is, for any \(g\in G\), there exists \(x_1,x_2,\ldots,x_m\in  \mathfrak{g}\) such that \(g=\exp x_1 \exp x_2 \cdots \exp x_m\)
  \end{enumerate}
\end{theorem}

\subsection{Resource theory of asymmetry}
RTA is a resource theory that treats the sensitivity of quantum states for the transformations caused by dynamics with symmetry.
We first define within RTA the free states and free operations, which are known as symmetric states and covariant operations, respectively.
We define them for a group \(G\) and its projective unitary representation \(\{U(g)\}_{g\in G}\) as follows \cite{Marvian_thesis}
\begin{definition}
	A symmetric state is a quantum state $\rho$ satisfying 
	\begin{equation}
		U(g) \rho U(g)^{-1}=\rho,\enskip\forall g\in G.
	\end{equation}
\end{definition}
Otherwise, the state is called asymmetric.
\begin{definition}
	A covariant operation is a completely positive and trace preserving (CPTP) map $\mathcal{E}$ satisfying
	\begin{equation}
		\mathcal{E} (U_{\mathrm{in}}(g) \rho U_{\mathrm{in}}(g)^{-1}) = U_{\mathrm{out}}(g) \mathcal{E} (\rho) U_{\mathrm{out}}(g)^{-1}
	\end{equation}
	for any $g\in G$ where \(\{U_{\mathrm{in}}(g)\}\) and \(\{U_\mathrm{out}(g)\}\) are  projective unitary representations of $G$ on input and output Hilbert spaces \(\mathcal{H}_{\mathrm{in}}\) and \(\mathcal{H}_{\mathrm{out}}\), respectively.
\end{definition}
From the above definition, we see that asymmetric states cannot be produced by symmetric states and covariant operations.
Indeed, if \(\mathcal{E}\) is a covariant operation, we know that, for any symmetric state \(\rho\), $\calE(\rho)$ is also a symmetric state:
\begin{align}
U_{\mathrm{out}}(g)\mathcal{E}(\rho)U_{\mathrm{out}}(g)^{-1}&=\mathcal{E}(U_{\mathrm{in}}(g) \rho U_{\mathrm{in}}(g)^{-1})\nonumber\\
& = \mathcal{E}(\rho),\enskip \forall g\in G.
\end{align}
For simplicity in the following discussion, we restrict ourselves to instances for which $\{U_{\mathrm{in}}(g)\}$ and $\{U_{\mathrm{out}}(g)\}$ are equal.

We here remark that the physical meaning of a covariant operation.
A unitary operation $V$ is said to be a covariant unitary if for any $g\in G$, $[V,U(g)]=0$.
Defining \(\mathcal{E}_V(\rho)\coloneqq V\rho V^{-1}\) for a covariant unitary operation $V$, $\mathcal{E}$ is a covariant operation.
In this case, \(\mathcal{E}_V\) is a quantum operation implemented by dynamics satisfying the conservation law.
The same is true for general covariant operations; specifically, the following theorem holds\cite{Marvian_thesis}.
\begin{theorem}
	Let $\{U(g)\}_{g\in G}$ be a projective unitary representation of a linear Lie group that describes a symmetry of system $S$ and a quantum operation $\mathcal{E}$ be a covariant operation on $S$.
	Then, there exists a Hilbert space $\mathcal{H}$ with a (non-projective) unitary representation $\{U_R(g)\}$ of $G$ and a symmetric pure state $\ket{\eta}$ in this space and a covariant unitary $V$ acting on Hilbert space $\mathcal{H}_S\otimes\mathcal{H}_R$ such that
	\begin{equation}
		\mathcal{E} (\rho) = \mathrm{tr} _R [V \rho\otimes \ketbra{\eta}{\eta} V^{-1}].
	\end{equation}
\end{theorem}
This theorem implies that a covariant operation is implemented by the dynamics that satisfies the conservation law and a state commuting with the conserved quantity.

\subsection{Fisher information}
When the group $G$ describing the symmetry is U(1), several known resource measures exist, in particular, the quantum Fisher information.
Although the quantum Fisher information for a quantum system depends on the inner product chosen, there is a well-known definition of  the quantum Fisher information applicable to a wide range of inner products \cite{MCFisher, Petz1996}.
We introduce this definition in the following.
A standard operator function $f$ is a monotonic function satisfying
\begin{enumerate}
  \item \(f(1)=1\)
  \item \(f(x)=xf(x^{-1})\)
  \item \(0 \le A \le B \Rightarrow f(A) \le f(B)\).
\end{enumerate}
Using this, we define the quantum Fisher information for a smooth quantum state family $\{\rho_t\}_{t\in\mathbb{R}}$ as 
\begin{equation}
	J^f_{\rho_t}=\mathrm{tr} \left[ \left( \frac{\partial \rho_t}{\partial t} \right) \left( \mathbb{J}^f_{\rho_t} \right) ^{-1} \left( \frac{\partial\rho_t}{\partial t} \right) \right]
\end{equation}
where \(\mathbb{J}^f_{\rho}=f(\mathbb{L_\rho}\mathbb{R_\rho}^{-1})\mathbb{R}_\rho\), \(\mathbb{L}_\rho(X)=\rho X\) and \(\mathbb{R}_\rho(X)=X\rho\).

A projective unitary representation of U(1) for system $S$ is written as \(U_t=e^{-itH}\) where $H$ is a periodic Hermitian operator ,i.e, \(e^{iH\tau}=1\) holds for some non-zero real number $\tau$.
The quantum Fisher information for a family of quantum states $\{\rho_t\}_{t\in\mathbb{R}}$ is
\begin{equation}
	F_\rho^f (H) = \left. J^f_{U_t \rho U_t^{-1}} \right|_{t=0},
\end{equation}
which is a resource measure satisfying Conditions 1--3 \cite{skew_resource,Takagi_skew}.
We derive this fact from the monotonicity of the quantum Fisher information under a CPTP map and the definition of the covariant operation.
Also, given the spectral decomposition of the quantum state $\rho$ with \(\rho=\sum_{i=1}^d p_i \ketbra{\psi_i}{\psi_i}\), we represent
\begin{equation}
    F_\rho^f(H)=\sum_{i,j=1}^d \frac{(p_i-p_j)^2}{p_j f \left(\frac{p_i}{p_j}\right)}\left|\Braket{\psi_i|H|\psi_j}\right|^2.
\end{equation}

\subsection{Relationship between Fisher information and quantum fluctuation}
The quantum Fisher information defined in the previous section has the property of a measure of quantum fluctuation.
Here we use "quantum fluctuation" as the part of fluctuation caused through quantum superposition.
To define what quantity can measure a quantum fluctuation,  we employ in this paper the criteria adopted by Luo \cite{Luo}.
Given a function $Q_\rho(H)$ of a quantum state $\rho$ and a Hermitian operator $H$, $Q_\rho(H)$ measures the quantum fluctuation of $H$ in $\rho$ if it satisfies Luo's criteria:
\begin{align}
    &0\le Q_{\rho}(H)\le V_{\rho}(H)\\
    &\rho\ is\ pure \Rightarrow Q_{\rho}(H)=V_{\rho}(H)\\
    &[\rho,H]=0 \Rightarrow  Q_\rho(H)=0\\
    &Q_{\sum_ip_i\rho_i}(H)\le\sum_ip_iQ_{\rho_i}(H)
\end{align}
where \(V_\rho(H)\) is the variance of $H$ in the state \(\rho\).

A metric adjusted skew information is defined as 
\begin{equation}
    I_{\rho}^f(H)=(f(0)/2)F_{\rho}^f(H)
\end{equation}
which satisfies Luo's criteria for any $f$ \cite{Hansen}.
Therefore, for any $f$, the quantum Fisher information is the quantum fluctuation of the conserved quantity.
Specifically, the quantum Fisher information for \(f(x)=(1+x)/2\) is called the symmetric logarithmic derivatve (SLD) Fisher information, which we write simply as \(F_{\rho}(H)\).
The SLD quantum Fisher information has excellent properties in describing quantum fluctuations, and concerning the variance the following theorem holds \cite{Marvian_distillation}.
\begin{theorem}\label{th:Marvian}
  Consider a system $S$ with a Hermitian operator $H_S$ and a quantum state $\rho$.
 The quantum Fisher information $F_\rho(H_S)$ can then be written as
  \begin{equation}
    F_{\rho}(H_S)=4\min_{\ket{\psi_{\rho}},H_R}V_{\ket{\psi_{\rho}}}(H_S+H_R)
  \end{equation}
  where \(V_{\ket{\phi}} (H) \) denotes the variance, which is defined as $V_{ \ket{\phi} } (H) := \braket{\phi|H^2|\phi} - \braket{\phi|H|\phi}^2$, and $\ket{\psi_\rho}$ is purification of the state $\rho$, and $H_R$ is a Hermitian operator of the auxiliary system $R$.
\end{theorem}
From this theorem, another critical theorem concerning the relationship between the quantum Fisher information and variance is derived \cite{min_V_Petz,min_V_Yu}.
\begin{theorem}\label{th:YU}
  Consider a system $S$ with a Hermitian operator $H_S$ and a quantum state $\rho$ with decomposition \(\rho=\sum_i p_i \ketbra{\phi_i}{\phi_i}\).
  Note that this decomposition is not necessarily a spectral decomposition and is not unique.
  Then, the quantum Fisher information $F_\rho(H_S)$ can be written as
  \begin{equation}
    F_\rho(H)=4\min_{p_i,\ket{\phi_i}}\sum_i p_i V_{\ket{\phi_i}}(H).
  \end{equation}
\end{theorem}

\section{Main results}\label{results}
In the previous section, we saw that Fisher information is a resource measure when U(1) represents the symmetry.
This section shows what quantity is a resource measure when the system has a more general symmetry than U(1).
As we shall see, the quantum Fisher information matrix is a resource measure for an arbitrary connected Lie group symmetry.
First, we consider the case where the $N$-dimensional real numbers $\mathbb{R}^N$ describe the symmetry. It corresponds to the situation where linearly independent multiple conserved quantities exist.
Next, we show that the quantum Fisher information matrix is a resource measure whenever a connected linear Lie group describes the symmetry.
We then see which properties of the quantum Fisher information when $G=U(1)$ carry over to the quantum Fisher information matrix.

\subsection{Fisher information matrix as a resource measure}
Let $\{\rho_t\}_{t\in \mathbb{R}^{N}}$  be a smooth quantum state family.
The $(i,j)$ component of the quantum Fisher matrix is defined as \cite{Petz2002}
\begin{equation}
	\left(J^f_{\rho_t}\right)_{ij}=\mathrm{tr} \left[ \left( \frac{\partial \rho_t}{\partial t_i} \right) \left( \mathbb{J}^f_{\rho_t} \right) ^{-1} \left( \frac{\partial\rho_t}{\partial t_j} \right) \right].
\end{equation}
Define the unitary operator $U_t=e^{-i\sum_k t_k X_k}$ with $X=(X_1,X_2,\ldots,X_N)$ as a set of $N$ linearly independent Hermitian operators.
The quantum Fisher information matrix for a family of quantum states $\{\rho_t\}_t=\{U_t\rho U_t^{-1}\}_t$ is denoted by
\begin{equation}
    \left( \hat{F}_\rho ^f (X_1,\ldots,X_N) \right)_{kl}=\left.\left( \hat{J}_{U_t\rho U_t^{-1}}^f \right)_{kl}\right|_{t=0}
\end{equation}
which is a function of the state $\rho$ and the set of operators $X_1,\ldots,X_N$.
In particular, given a quantum state \(\rho\) with spectral decomposition \(\rho=\sum_{i=1}^d p_i\ketbra{\psi_i}{\psi_i}\), the quantum Fisher information matrix can be expressed  as
\begin{equation}
    \left( \hat{F}_{\rho}^f(\{X_k\}) \right)_{kl}=\sum_{i,j=1}^d\frac{(p_i-p_j)^2}{p_j f\left( \frac{p_i}{p_j} \right)} \Braket{\psi_i|X_k|\psi_j}\Braket{\psi_j|X_l|\psi_i}
\end{equation}
The following theorem implies that the quantum Fisher information matrix $\hat{F}_\rho^f(\{X_n\}^{N}_{n=1})$ is a resource measure when there are multiple conserved quantities.
\begin{theorem}\label{th:multiple-conserved}
  Let $(X_1,X_2,\ldots,X_N)$ be a set of linearly independent Hermitian operators and $U_t=e^{-i\sum_k t_k X_k}$ be a projective unitary representation of the $N$-dimensional real numbers $\mathbb{R}^N$.
  Then, the quantum Fisher information matrix is a resource measure, i.e., it satisfies
  \begin{align}
      &\hat{F}_{\rho}^f(X_1,\ldots,X_N)\ge0 \label{Fisher-matrix-positivity} \\
      &\hat{F}_{\rho}^f(X_1,\ldots,X_N)=0\Leftrightarrow [\rho,U_t]=0,\ \forall t\in \mathbb{R}^N \label{Fisher-matrix-free-state} \\
      &\hat{F}_\rho^f(X_1,\ldots,X_N)\ge\hat{F}_{\mathcal{E}(\rho)}^f(X_1,\ldots,X_N) \label{Fisher-matrix-free-operation}
  \end{align}
  where \(\mathcal{E}\) is a covariant operation. Here, all of the above inequalities are matrix inequalities.
\end{theorem}

We can extend the above theorem to the more general case, in which a more general Lie group $G$ describes the symmetry.
\begin{theorem}\label{th:resoruce-measure1}
  Let $G$ be a linear Lie group and $\{U(g)\}_{g\in G}$ be a projective unitary representation of $G$ on a Hilbert space $\mathcal{H}$.
  Then, there exists a map \(\varphi\) from the Lie ring $\mathfrak{g}$ of $G$ to the Lie ring $\mathfrak{u}(\mathcal{H})$ such that
  \begin{equation}
      \hat{F}_\rho^f(\{X_n\})=0\Rightarrow [\rho,U(g)]=0,\quad \forall g\in G\label{eq:fmrsp}
  \end{equation}
  where \(X_l\coloneqq -i\varphi(x_l)\) with basis $\{x_l\}$ in $\mathfrak{g}$.
  If $G$ is connected, the converse is also true.
  Furthermore, the quantum Fisher information matrix $\hat{F}_\rho^f(\{X_n\})$ is monotonic under a covariant operation, i.e.
   \begin{equation}\label{eq:mononicity-general}
      \hat{F}_\rho^f(\{X_l\})\ge{F}_{\mathcal{E}(\rho)}^f(\{X_l\})
  \end{equation}
  where $\mathcal{E}$ is covariant operation.
\end{theorem}
This theorem shows that when $G$ is a connected Lie group, the quantum Fisher information for any $f$ is a resource measure satisfying Conditions 1--3. 

In addition to its property as a resource measure, the quantum Fisher information matrix has the following property called selective monotonicity.
\begin{theorem}\label{th:selective}
  Let $G$ be a linear Lie group, $\{U(g)\}_{g\in G}$ be a projective unitary representation of $G$ on  Hilbert space $\mathcal{H}$, and $\mathcal{E}=\sum_j \mathcal{E}_j$ be a quantum operation with a covariant CP trace non-increasing map $\mathcal{E}_j$.
  The quantum Fisher information does not increase on average, i.e.,
  \begin{equation}
      \hat{F}_\rho^f(\{X_n\})\ge \sum_j p_j \hat{F}_{\sigma_j}^f(\{X_n\})
  \end{equation}
  where $p_j=\mathrm{tr}[\mathcal{E}_j(\rho)]$ and $\sigma_j=\mathcal{E}_j(\rho)/p_j$.
\end{theorem}

\subsection{Relationship between quantum Fisher information matrix and quantum fluctuation}
When there is only one conserved quantity, the quantum Fisher information satisfies Luo's criteria and the physical meaning of quantifying the quantum fluctuation.
In considering the physical meaning of the quantum Fisher information matrix, we first show that Luo's criteria can be extended to the form applicable to matrices and that the quantum Fisher information matrix satisfies the criteria.
Specifically, the following theorem holds.
\begin{theorem}\label{th:q-fluc}
  Defining $\hat{I}^f_\rho(\{X_n\}\coloneqq\frac{f(0)}{2}\hat{F}_\rho^f(\{X_n\})$, it satisfies
  \begin{align}
    &0\le \hat{I}_{\rho}^f(\{X_n\})\le \hat{V}_{\rho}(\{X_n\})\label{eq:q-fluc1}\\
    &\rho\ is\ pure \Rightarrow \hat{I}_{\rho}^f(\{X_n\})=\hat{V}_{\rho}^f(\{X_n\})\label{eq:q-fluc2}\\
    &[\rho,X_k]=0,\ \forall k \Rightarrow  \hat{I}_\rho^f(\{X_n\})=0\label{eq:q-fluc3}\\
    &\hat{I}_{\sum_i p_i\rho_i}^f(\{X_n\})\le\sum_i p_i \hat{I}_{\rho_i}^f(\{X_n\}).\label{eq:q-fluc4}
\end{align}
\end{theorem}
This theorem shows that the quantum Fisher information matrix for any $f$ is a measure of quantum fluctuation. 

Furthermore, let us consider multivariable extensions of Theorem \ref{th:Marvian} and Theorem \ref{th:YU}. First, the following theorem holds as a multivariable extension of Theorem \ref{th:Marvian}.
\begin{theorem}\label{th:min-V}
  Consider a system $S$ with a set of linearly independent Hermitian operators $\{X_n^S\}$ and a quantum state $\rho$.
  The quantum Fisher information matrix $\hat{F}_\rho(\{X_n^F\})$ can be written as
  \begin{equation}
    \hat{F}_{\rho}(\{X_n\}) = 4\min_{\ket{\psi_{\rho}},\{X_n^R\}}\hat{V}_{\ket{\psi_\rho}}(\{X_n^S+X_n^R\})
  \end{equation}
   where \(\hat{V}_{\ket{\phi}} (H) \) is a covariance matrix, defined as $(\hat{V}_{\rho}(\{X_n\}))_{kl}=\frac{1}{2}\langle (X_k-\langle X_k\rangle_{\rho} ) (X_l- \langle X_l \rangle_{\rho} )	+ (X_l-\langle X_l \rangle_{\rho} ) (X_k-\langle X_k\rangle_{\rho} ) \rangle_{\rho}$  and $\langle Y\rangle_{\sigma}$ is expected value of a Hermitian operator $Y$ in a state $\sigma$, defined as $\langle Y\rangle_{\sigma}\coloneqq\Tr[\sigma Y]$ and $\ket{\psi_\rho}$ is purification of the state $\rho$ and $\{X_n^R\}$ is a set of Hermitian operators of auxiliary system R.
\end{theorem}
This theorem is proved by giving another proof of Theorem \ref{th:Marvian} and considering its multivariable extension.
In the proof, we combine Uhlmann's theorem that that the fidelity of the mixed state is equal to the maximum of the fidelity of the purification and the fact that the second-order derivative of fidelity is equal to the SLD Fisher information.

Finally, we show that, by constructing a counterexample, the multivariable extension of Theorem \ref{th:YU} does not hold.
The counterexample is as follows. Consider a qubit system $S$ with state \(\rho=I/2\) with $I$ the identity.
For noncommutative Hermitian operators \(X_1\) and \(X_2\) are, we have
\begin{equation}
  \hat{F}_\rho (X_1, X_2)=0.
\end{equation}
Because \(X_1\) and \(X_2\) are noncommutative, for any decomposition \(\rho=\sum_i p_i \ketbra{\phi_i}{\phi_i}\), either \(V_{\ket{\phi_i}}(X_1)>0\) or \(V_{\ket{\phi_i}}(X_2)>0\) holds.
This implies
\begin{equation}
  \sum_ip_i\hat{V}_{\ket{\phi_i}} (X_1,X_2)>0.
\end{equation}
Hence, the multivariable extension of Theorem \ref{th:YU} does not hold.

\section{Conclusion}\label{conclusion}
We have shown that the quantum Fisher matrix is a resource measure when a connected linear Lie group describes the symmetry, i.e., the following conditions are satisfied.
\begin{itemize}
\item{\(\hat{F}_\rho(\{X_k\})\ge 0\)}
\item{\(\hat{F}_\rho(\{X_k\})=0\Leftrightarrow \rho\mbox{ is a symmetric state}\)}
\item{If \(\mathcal{E}\) is a covariant operation, \(\hat{F}_\rho(\{X_k\})\ge \hat{F}_{\mathcal{E}(\rho)}(\{X_k\})\) holds.}
\end{itemize}
In proving that the quantum Fisher information matrix satisfies these conditions, we used the fact that a connected linear Lie group is generated by its Lie ring and the properties of the representation of the Lie ring obtained from its unitary or projective unitary representations.
Our result is a generalization of the quantum Fisher information as a resource measure with U(1) symmetry and is the first example of a resource measure being a matrix in resource theory.

We also extended three known results, given as theorems, concerning the physical meaning of the quantum Fisher information for multiple conserved quantities.
The first is Luo's criteria, which we extended to be applicable to matrices and showed that the quantum Fisher information matrix satisfies these extended criteria.
The second is that the quantum Fisher information matrix is equal to the minimum of the covariance matrix of its purification, and we gave another proof using Ullman's theorem.
As an extension of this theorem for multiple conserved quantities, we showed that for the quantum Fisher information matrix the equality
\begin{equation}
    \hat{F}_{\rho}(\{X_n\}) =4\times\min_{\ket{\psi_{\rho}},\{X_n^R\}}\hat{V}_{\ket{\psi_\rho}}(\{X_n^S+X_n^R\})
\end{equation}
holds.
The third theorem states that the quantum Fisher information is equal to the minimum average of the variance of the pure state. And we showed, by giving a counterexample, that this theorem cannot be extended to general symmetry scenario.

\textbf{Note added:}
After completing our analysis, a new paper \cite{Marvian_new} was uploaded on arXiv. The paper contains another proof of Theorem \ref{th:Marvian} for U(1) symmetry.
The proof uses a similar method of the proof for Theorem \ref{th:min-V}, although Theorem \ref{th:min-V} treats general symmetries.
We also remark that the main results of our article are features of the quantum Fisher information matrices for the case of the general symmetry, and the main results have no overlap with Ref. \cite{Marvian_new}.

\begin{acknowledgments}
	The present work was supported by JSPS Grants-in-Aid for Scientific Research No. JP19K14610 (HT), JST PRESTO No. JPMJPR2014 (HT), and  JST MOONSHOT (H. T. Grant
No. JPMJMS2061). 
We thank Richard Haase, PhD, from Edanz for editing a draft of this manuscript. 
\end{acknowledgments}

\bibliographystyle{apsrev4-2}

\appendix
\section{Proof of Theorem \ref{th:multiple-conserved}}
  For any \(\lambda \in \mathbb{R}^N\), we have 
  \begin{align}
    &\lambda^T\hat{F}_{\rho}^f(X_1,\ldots,X_N)\lambda\notag\\
    &=\sum_{i,j=1}^d\sum_{k,l=1}^N\lambda_k\lambda_l\frac{(p_i-p_j)^2}{p_j f\left(\frac{p_i}{p_j}\right)}\Braket{\psi_i|X_k|\psi_j}\Braket{\psi_i|X_l|\psi_i}\nonumber\\
    &=\sum_{i,j=1}^d\frac{(p_i-p_j)^2}{p_j f\left(\frac{p_i}{p_j}\right)}\left|\Braket{\psi_i|\sum_{k=1}^N\lambda_kX_k|\psi_j}\right|^2\nonumber\\
    &=F_{\rho}^f\left(\sum_{k=1}^N\lambda_k X_k\right).\label{Fisher-matrix-and-Fisher-information}
  \end{align}
  Therefore, we have \eqref{Fisher-matrix-positivity} from positivity of the Fisher information.
  Then, note that
  \begin{align}
    &\hat{F}_{\rho}^f(X_1,\ldots,X_N)=0\notag\\
    &\Longleftrightarrow\lambda^T\hat{F}_{\rho}^f (X_1,\ldots,X_N)\lambda=0,\ \forall \lambda\in\mathbb{R}^N\nonumber\\
    &\Longleftrightarrow F_{\rho}^f \left(\sum_{k=1}^N\lambda_k X_k\right)=0,\ \forall \lambda\in\mathbb{R}^N\nonumber\\
    &\Longleftrightarrow\left[e^{-it\sum_{k=1}^N\lambda_k X_k},\rho\right]=0,\ \forall \lambda \in \mathbb{R}^N,\ t\in\mathbb{R},
  \end{align}
  which implies \eqref{Fisher-matrix-free-state}.
  Given the monotonicity of the quantum Fisher information under a covariant operation, we have 
  \begin{align}
    \lambda^T\hat{F}_{\rho}^f (X_1,\ldots,X_N)\lambda&=F_{\rho}^f \left(\sum_k\lambda_kX_k \right)\nonumber\\
    &\ge F_{\mathcal{E}(\rho)}^f \left(\sum_k\lambda_kX_k\right)\nonumber\\
    &=\lambda^T\hat{F}_{\mathcal{E}(\rho)}^f (X_1,\ldots,X_N)\lambda
  \end{align}
  which implies \eqref{Fisher-matrix-free-operation}.

\section{Proof of theorem \ref{th:resoruce-measure1}}
  Let \(\overline{\mathcal{H}}\) be a Hilbert space of the same dimension as $\mathcal{H}$.
  We consider
  \begin{equation}
      X=\sum_{i,j}x_{i,j}\ketbra{i}{j}_{\mathcal{H}}\ \leftrightarrow\ \ket{X}\rangle_{\mathcal{H}\otimes\overline{\mathcal{H}}}=\sum_{i,j} x_{i,j}\ket{i}_{\mathcal{H}}\otimes\ket{j}_{\overline{\mathcal{H}}}.
  \end{equation}
  Thus, we find
  \begin{equation}
      Y\otimes Z\ket{X}\rangle=\ket{YXZ^T}\rangle
  \end{equation}
  where $X^T$ is transposed of $X$ for basis \(\{\ket{j}_{\overline{\mathcal{H}}}\}\).
  For any $g\in G$ and $\rho \in \mathcal{H}$ we find
  \begin{equation}
      U_g\otimes (U_g)^\ast \ket{\rho}\rangle=\ket{U_g\rho U_{g}^\dagger}\rangle
  \end{equation}
  where $X^\ast$ denotes the complex conjugate of $X$ for basis \(\{\ket{j}_{\overline{\mathcal{H}}}\}\).
  
  Here, we define \(\Phi\colon G\to U(\mathcal{H}\otimes\overline{\mathcal{H}})\) as
  \begin{equation}
      \Phi(g)\coloneqq U_g\otimes(U_g)^\ast.
  \end{equation}
  Then, we have
  \begin{align}
      \Phi(g)\Phi(g')&=(U_g\otimes U_g^\ast)(U_{g'}\otimes U_{g'}^\ast)\nonumber\\
      &=U_g U_{g'}\otimes U_g^\ast U_{g'}^\ast\nonumber\\
      &=U_g U_{g'}\otimes (U_g U_{g'})^\ast\nonumber\\
      &=e^{i\omega(g,g')}U_{g g'}\otimes (e^{i\omega(g,g')}U_{g g'})^\ast\nonumber\\
      &=U_{g g'}\otimes U_{g g'} ^\ast.
  \end{align}
  This implies $\Phi$ is a homomorphism. 
  Therefore, using Theorem \ref{Th-differential-representation}, there exists a unique linear map \(\tilde{\varphi}\colon \mathfrak{g}\to L(U(\mathcal{H}\otimes\overline{\mathcal{H}}))\) such that
  \begin{equation}
      \forall x\in \mathfrak{g},\quad e^{\tilde{\varphi}(x)}=\Phi(e^x).
  \end{equation}
   To identify the form of \(\tilde{\varphi}(x)\), we define
  \begin{equation}
      \tilde{\tilde{\varphi}}(x)\coloneqq(\log U_{e^x})\otimes I_{\overline{\mathcal{H}}}+I_{\mathcal{H}}\otimes(\log U_{e^x})^\ast,
  \end{equation}
  where \(I_{\mathcal{H}}\) and \(I_{\overline{\mathcal{H}}}\) denote the identity operators of \(\mathcal{H}\) and \(\overline{\mathcal{H}}\), respectively.
  Then, the map \(\tilde{\tilde{\varphi}}\) satisfies
  \begin{equation}
      \forall x\in \mathfrak{g} ,\quad e^{\tilde{\tilde{\varphi}}(x)}=\Phi(e^x).
  \end{equation}
  From the uniqueness of \(\tilde{\varphi}\), we have \(\tilde{\varphi}=\tilde{\tilde{\varphi}}\).
  Using this map \(\tilde{\varphi}\), we can define \(\varphi\) noting that \(\tilde{\varphi}\) can be written as
  \begin{equation}
      \tilde{\varphi}=Y(x)\otimes I_{\overline{\mathcal{H}}}+I_{\mathcal{H}}\otimes Y(x)^\ast,
  \end{equation}
  where $Y(x)$ denotes a skew-Hermitian operator.
  It follows from this form of $\tilde{\varphi(x)}$ that \(Y(x)\) is defined up to a constant multiplication of \(I_{\mathcal{H}}\).
  Let $Y$ and $Y'$ be skew-Hermitian operators. 
  If  \(Y\otimes I_{\overline{\mathcal{H}}}+I_{\mathcal{H}}\otimes Y^\ast=Y'\otimes I_{\overline{\mathcal{H}}}+I_{\mathcal{H}}\otimes Y^{\prime\ast}\), we have
  \begin{equation}
      (Y-Y')\otimes I_{\overline{\mathcal{H}}}=I_{\mathcal{H}}\otimes (Y'-Y)^\ast
  \end{equation}
  and for any operator $A$ on $\mathcal{H}$, 
  \begin{equation}
      [(Y-Y')\otimes I_{\overline{\mathcal{H}}},I_{\mathcal{H}}\otimes A]=0.
  \end{equation}
  From these two equations, we have
  \begin{equation}
      [(Y-Y')^\ast,A]=0.
  \end{equation}
  Because \(Y\) is a skew-Hermitian operator and the above commutation relation holds for any \(A\), there exists a real number \(c\) such that 
  \begin{equation}
      Y'-Y=icI.
  \end{equation}
  
  We define map  \(\varphi\colon\mathfrak{g} \to \mathfrak{u}(\mathcal{H})\) to be
  \begin{equation}
      \varphi(x)\coloneqq Y(x).
  \end{equation}
  This \(\varphi\) is not necessarily linear. However, from the linearity of \(\tilde{\varphi}\), $\varphi$ satisfies
  \begin{align}
      \forall a\in \mathbb{R},\forall x\in \mathfrak{g},\exists b\in \mathbb{R},\quad \mathrm{s.t.}\quad \varphi(ax)=a\varphi(x)+i b I_{\mathcal{H}} \label{eq:pdll1} \\
      \forall x,y\in \mathfrak{g},\exists c\in \mathbb{R},\quad\mathrm{s.t.}\quad \varphi(x+y)=\varphi(x)+\varphi(y)+i c I_{\mathcal{H}}\label{eq:pdll2}.
  \end{align}
  Defining \(\varphi\) as above, and then defining \(X_l\coloneqq-i\varphi(x_l)\) for the basis of \(\mathfrak{g}\), we have
  \begin{align}
      \hat{F}_\rho^f(\{X_l\})=0&\Leftrightarrow [\rho, e^{i\sum_k \lambda_k X_k}]=0,\quad\forall\lambda\in \mathbb{R}^N\nonumber\\
      &\Leftrightarrow [\rho,e^{\varphi(x)}]=0,\quad\forall x\in \mathfrak{g}\nonumber\\
      &\Leftrightarrow e^{\tilde{\varphi}(x)}\ket{\rho}\rangle=\ket{\rho}\rangle,\quad\forall x\in \mathfrak{g}\label{eq:fmzep}
  \end{align}
  where, to establish the second line, we have used \eqref{eq:pdll1}, \eqref{eq:pdll2}, and \(x=\sum_l\lambda_l x_l\) for any $x\in \mathfrak{g}$; to obtain the last line, we used \(e^{\tilde{\varphi}}=e^{\varphi(x)}\otimes e^{\varphi(x)^\ast}\).
  
  We show \eqref{eq:fmrsp} from \eqref{eq:fmzep}.
  Assuming that \([\rho,U_g]=0\) for any \(g\in \mathfrak{g}\), we have
  \begin{equation}
      \Phi(e^x)\ket{\rho}=\ket{\rho}\rangle.
  \end{equation}
  From \(\Phi(e^x)=e^{\tilde{\varphi}(x)}\), we have
  \begin{equation}
      e^{\tilde{\varphi}(x)}\ket{\rho}\rangle=\ket{\rho}\rangle.
  \end{equation}
  Using (\ref{eq:fmzep}), this implies \(\hat{F}_\rho^f(\{X_l\})=0\).
  
  Next, we show that the converse holds if \(G\) is a connected.
  Given Theorem \ref{Th-connected}, if $G$ is connected, for any \(g\in G\) there exists \(y_1,\ldots,y_m\in \mathfrak{g}\) such that
  \begin{equation}
      g=e^{y_1}\cdots e^{y_m}.
  \end{equation}
  Therefore,
  \begin{align}
      \Phi(g)&=\Phi(e^{y_1}\cdots e^{y_m})\nonumber\\
      &=\Phi(e^{y_1})\cdots \Phi(e^{y_m})\nonumber\\
      &=e^{\tilde{\varphi}(y_1)}\cdots e^{\tilde{\varphi}(y_m)}.
  \end{align}
   Using (\ref{eq:fmzep}), this implies \(\Phi(g)\ket{\rho}\rangle=\ket{\rho}\rangle\) if $\hat{F}_\rho^f(\{X_l\})=0$.
   Therefore, \([\rho,U_g]=0\) holds.
  
  Next, we prove Theorem \eqref{eq:mononicity-general}.
  For any vector $\lambda\in \mathbb{R}^N$, there exists a real number $c$ such that 
  \begin{equation}
      \varphi\left( \sum_k \lambda_k x_k \right)=-i\sum_k \lambda_k  X_k+icI.   
  \end{equation}
  Thus, we have
  \begin{equation}
      e^{\varphi(\sum_k\lambda_k x_k)}=e^{ic}e^{-i\sum_k \lambda_k X_k}.
  \end{equation}
  From \(U_{e^x}\rho U_{e^x}^{-1}=e^{\varphi(x)}\rho e^{-\varphi(x)}\), we obtain
  \begin{align}
      &e^{-i\sum_k \lambda_k X_k}\mathcal{E}(\rho)e^{i\sum_k \lambda_k X_k}\notag\\
      &=e^{\varphi(\sum_k \lambda_k x_k)}\mathcal{E}(\rho)e^{\varphi(\sum_k \lambda_k x_k)}\notag\\
      &=\mathcal{E}(e^{\varphi(\sum_k \lambda_k x_k)}\rho e^{-\varphi(\sum_k \lambda_k x_k)})\notag\\
      &=\mathcal{E}(e^{-i\sum_k \lambda_k X_k}\rho e^{i\sum_k \lambda_k X_k})\label{eq:eEe}
  \end{align}
  Using Theorem \ref{th:multiple-conserved}, we have
  \begin{equation}
      \hat{F}_\rho^f(\{X_l\})\ge\hat{F}_{\mathcal{E}(\rho)}^f(\{X_l\}).
  \end{equation}

\section{Proof of Theorem \ref{th:selective}}
  Considering $\varphi$ and $\{X_n\}$ to be the same as those defined in Theorem \ref{th:resoruce-measure1}, we have in similar manner to \eqref{eq:eEe},
  \begin{equation}
      \mathcal{E}_j(e^{-i\sum_k \lambda_k X_k}\rho e^{-\sum_k \lambda_k X_k})=e^{-i\sum_k \lambda_k X_k}\mathcal{E}_j(\rho)e^{i\sum_k \lambda_k X_k}.
  \end{equation}
  With the selective monotonicity of Fisher information, this implies
  \begin{equation}
      F_\rho^f(\sum_k \lambda_k X_k)\ge \sum_j p_j F_\rho^f(\sum_k \lambda_k X_k)
  \end{equation}
  Using \eqref{Fisher-matrix-and-Fisher-information}, we have
  \begin{equation}
      \hat{F}_\rho^f(\{X_n\})\ge \sum_j p_j F_\rho^f(\{X_n\}).
  \end{equation}
  
\section{Proof of Theorem \ref{th:q-fluc}}
  With the quantum Fisher information satisfying Luo's criteria, for any $\lambda\in \mathbb{R}^N$, we have
  \begin{align}
    &0\le I_{\rho}^f(\sum_k \lambda_k X_k)\le V_{\rho}(\sum_k \lambda_k X_k)\label{eq:q-fluc1-1}\\
    &\rho\ is\ pure \Rightarrow I_{\rho}^f(\sum_k \lambda_k X_k)=V_{\rho}(\sum _k \lambda_k X_k)\\
    &[\rho,X_l]=0,\ \forall l \in \{1,\ldots,N\} \Rightarrow  I_\rho^f(\sum_k \lambda_k X_k)=0\\
    &I_{\sum_ip_i\rho_i}^f(\sum_k \lambda_k X_k)\le\sum_ip_iI_{\rho_i}^f(\sum_k \lambda_k X_k).
  \end{align}
  Note that
  \begin{align}
      &\lambda^T\hat{V}_{\rho}(\{X_n\})\lambda\nonumber\\
      &=\left\langle \sum_{k,l} \frac{\lambda_k\lambda_l+\lambda_l\lambda_k}{2} \left(X_k-\left\langle X_k\right\rangle_{\rho}\right)\left(X_l-\left\langle X_l\right\rangle_{\rho}\right)\right\rangle_{\rho}\nonumber\\
      &=\left\langle \left(\sum_k\lambda_k \left(X_k-\left\langle X_k\right\rangle_{\rho}\right)\right) \left(\sum_l\lambda_l\left(X_l-\left\langle X_l\right\rangle_{\rho}\right)\right)\right\rangle_{\rho}\nonumber\\
      &=\left\langle \left( \sum_k\lambda_k X_k-\left\langle \sum_k\lambda_k X_k \right\rangle_{\rho} \right)^2\right\rangle_{\rho}\nonumber\\
      &=V_\rho \left( \sum_k \lambda_k X_k \right).\label{eq:lVl}
  \end{align}
  Using the above equation and \eqref{Fisher-matrix-and-Fisher-information}, we rewrite \eqref{eq:q-fluc1-1} as
  \begin{equation}
      0\le\lambda^T\hat{I}_\rho^f(\{X_n\})\lambda\le\lambda^T\hat{V}_\rho(\{X_n\})\lambda,
  \end{equation}
  which implies \eqref{eq:q-fluc1}.
  Similarly, we have \eqref{eq:q-fluc2}\eqref{eq:q-fluc3}\eqref{eq:q-fluc4}.
  
\section{Proof of Theorem \ref{th:min-V}}
  First, we provide a proof for the one-variable case different from Ref \cite{Marvian_distillation}.
  Let \(\ket{\psi_\rho}\) be any purification of \(\rho\), and let \(X^S\) and \(X^R\) be any Hermitian operator of systems $S$ and $R$.
  We define \(\rho_t\coloneqq e^{-it X^S}\rho e^{it X^S}\) and its purification $\ket{\psi_{\rho_t}}$ such that
  \begin{equation}
      \ket{\psi_{\rho_t}}\coloneqq e^{-it X^R}\otimes e^{-it X^S}\ket{\psi_\rho}.
  \end{equation}
  From Uhlmann's theorem, we have
  \begin{equation}
      F(\rho,\rho_t) \ge F(\ket{\psi_\rho},\ket{\psi_{\rho_t}}).
  \end{equation}
  Because the second-order derivative of fidelity is equal to the quantum Fisher information, this implies
  \begin{align}
      F_\rho(X^S)&=8\lim_{t\to 0} \frac{1-F(\rho,\rho_t)}{t^2}\nonumber\\
      &\le8\lim_{t\to 0}\frac{1-F(\ket{\psi_\rho},\ket{\psi_{\rho_t}}}{t^2}\nonumber\\
      &=F_{\ket{\psi_\rho}}(X^S+X^R)\nonumber\\
      &=4V_{\ket{\psi_\rho}}(X^S+X^R).
  \end{align}
  We show that there exists purification \(\ket{\psi_\rho}\) and \(X^R\) that achieves the equality of the above inequality.
  From Uhlmann's theorem, there exists purification \(\ket{\psi_\rho}\) and \(\ket{\psi_{\rho_t}}\) such that
  \begin{equation}
      F(\rho,\rho_t)=\left| \Braket{\psi_\rho|\psi_{\rho_t}} \right|.
  \end{equation}
  These states are given by
  \begin{align}
      \ket{\psi _{\rho}}=I_R\otimes\sqrt{\rho}\ket{\Phi _+}\\
      \ket{\psi _{\rho _t}}=I_R\otimes\sqrt{\rho _t}V_t ^{-1}\ket{\Phi _+}
  \end{align}
  where \(\ket{\Phi _+}\coloneqq\sum _i \ket{i}\ket{i}\) with any basis \(\{\ket{i}\}\) and the Hilbert spaces of $S$ and $R$ are considered to be the same.
  The operator $V_t$ is a unitary operator given by a polar decomposition of \(\sqrt{\rho}\sqrt{\rho _t}\) such that
  \begin{equation}
      \sqrt{\rho}\sqrt{\rho _t}=V_t\sqrt{\sqrt{\rho _t}\rho\sqrt{\rho _t}}
  \end{equation}
  Given the spectral decomposition of the quantum state \(\rho\) with \(\rho=\sum_{i=1}^d p_i\ketbra{\varphi_i}{\varphi_i}\), we have
  \begin{equation}
      \sqrt{\rho _t}=U_t\sqrt{\rho}U_t^{-1}.
  \end{equation}
  Note that for any operator $X$
  \begin{equation}
      I_R\otimes X\ket{\Phi_+}=X^T\otimes I_S\ket{\Phi_+}.
  \end{equation}
  Using above equation, we describe
  \begin{equation}
      \ket{\psi _{\rho _t}}=\left(U_t^{-1}V_t^{-1}\right)^T\otimes U_t\sqrt{\rho}\ket{\Phi_+}=W_t\otimes U_t\ket{\psi _{\rho}}
  \end{equation}
  where \(W_t:=\left(U_t^{-1}V_t^{-1}\right)^T\).
  Because $U_t$ and $V_t$ are unitary operators, $W_t$ is also a unitary operator, which we express $W_t$ as
  \begin{equation}
      W_t=I+tA+t^2B+O(t^3),
  \end{equation}
  which leads to
  \begin{equation}
      W_t^{\dagger} W_t=I+t(A+A^{\dagger})+t^2(B+B^{\dagger}+A^{\dagger}A)+O(t^3).
  \end{equation}
  Because \(W_t\) is a unitary operator, we have 
  \begin{gather}
      A+A^\dagger=0\\
      B+B^{\dagger}+A^{\dagger}A=0.
  \end{gather}
  From the first equation, $A$ is skew-Hermitian, i.e., there exists a Hermitian operator $X^R$ such that $A=-i X^R$.
  From the second equation, we describe 
  \begin{equation}
      B=-\frac{(X^R)^2}{2}+iC
  \end{equation}
  where $C$ is a Hermitian operator.
  Therefore, we express \(W_t\) as
  \begin{equation}
      W_t=I-it X^R-t^2\left(\frac{(X^R)^2}{2}-iC\right)+O(t^3).
  \end{equation}
  Expanding the fidelity to second order in \(t\), we obtain
  \begin{align}
      &F(\rho,\rho_t)\nonumber\\
      &=|\Braket{\psi_{\rho}|W_t\otimes U_t|\psi_\rho}|\nonumber\\
      &=\sqrt{1-t^2V_{\ket{\psi_{\rho}}}(X^S+X^R)+O(t^3)}\nonumber\\
      &=1-\frac{t^2}{2}V_{\ket{\psi_{\rho}}}(X^S+X^R)+O(t^3)
  \end{align}
  Therefore, because the second-order derivative of fidelity is equal to the Fisher information, we have
  \begin{equation}
      F_\rho(X^S)=8\lim_{t\to 0}\frac{1-F(\rho,\rho_t)}{t^2}=4V_{\ket{\psi_\rho}}(X^S+X^R)
  \end{equation}
  Next, we show instances with multiple variable.
  Let \(\lambda\in\mathbb{R}^N\) be any real vector and \(\{X_n\}\) be a set of Hermitian operators.
  Using \eqref{Fisher-matrix-and-Fisher-information} and \eqref{eq:lVl} and the equality of the quantum Fisher information for a pure state and the variance, then for any pure state $\ket{\psi}$, we have
  \begin{align}
      \lambda^T\hat{F}_{\ket{\psi}}(\{X_n\})\lambda&=F_{\ket{\psi}}\left(\sum_k\lambda_k X_k\right)\nonumber\\
      &=4\times V_{\ket{\psi}}\left( \sum_k\lambda_k X_k \right)\nonumber\\
      &=4\times \lambda^T\hat{V}_{\ket{\psi}}(\{X_n\})\lambda
  \end{align}
  which implies
  \begin{equation}
      \hat{F}_{\ket{\psi}}(\{X_n\})=4\hat{V}_{\ket{\psi}}(\{X_n\}).
  \end{equation}
  That is, the quantum Fisher information matrix for a pure state is equivalent to a covariance matrix.
  Let $\{X_n^R\}$ be a set of Hermitian operators of the auxiliary system $R$.
  We define the state $\rho_{t\lambda}$ as
  \begin{equation}
      \rho_{t\lambda}=e^{-it\sum_k\lambda_kX_k^S}\rho e^{it\sum_k\lambda_kX_k^S}=U_{t\lambda}\rho U_{t\lambda}^{-1}.
  \end{equation}
  Then, we define the purification of $\rho_{t\lambda}$ as
  \begin{equation}
      \ket{\psi_{\rho_{t\lambda}}}=e^{-it\sum_k\lambda_kX_k^R}\otimes e^{-it\sum_k\lambda_k X_k^S}\ket{\psi_{\rho}},
  \end{equation}
  where \(\ket{\psi_\rho}\) is the purification of $\rho$.
  Using Ulmann's theorem and the fact that the second-order derivative of fidelity is equal to the quantum Fisher information, we have
  \begin{align}
      \lambda^T\hat{F}(\{X_n^S\})\lambda&=F_{\rho}\left( \sum_k\lambda_kX_k^S \right)\nonumber\\
      &=8\lim_{t\to 0}\frac{1- F(\rho,\rho_{t\lambda})}{t^2}\nonumber\\
      &\le 8\lim_{t\to 0}\frac{1-F(\ket{\psi_{\rho}} ,\ket{\psi_{\rho_{t\lambda}}})}{t^2}\nonumber\\
      &F_{\ket{\psi_\rho} }\left( \sum_k\lambda_k(X_k^S+X_k^R) \right)\nonumber\\
      &=\lambda^T\hat{F}_{\ket{\psi_{\rho}}}\left(\{X_n^S+X_n^R\}\right)\lambda
  \end{align}
  which implies
  \begin{equation}
  \begin{split}
      \hat{F}_{\rho}(\{X_n^S+X_n^R\})&\le\hat{F}_{\ket{\psi_\rho}}(\{X^S_n+X^R_n\})\\
      &=4\hat{V}_{\ket{\psi_\rho}}(\{X^S_n+X^R_n\}).
  \end{split}  
  \end{equation}
  We show that there exists purification \(\ket{\psi_\rho}\) and \(X^R\) that provides the equality of the above inequality.
  Applying Uhlmann's theorem, there exist states \(\ket{\psi_\rho}\) and \(\ket{\psi_{\rho_{t\lambda}}}\) such that
  \begin{equation}
      F(\rho,\rho_{t\lambda})=\left|\Braket{\psi_\rho
      |\psi_{\rho_{t\lambda}}}\right|.
  \end{equation}
   Similar to single conserved quantity cases, these states are given by
  \begin{align}
      &\ket{\psi_{\rho}}=I_R\otimes \sqrt{\rho}\ket{\Phi_+}\\
      &\ket{\psi_{\rho_{t\lambda}}}=I_R\otimes \sqrt{\rho_{t\lambda}}V_{t\lambda}^{\dagger}\ket{\Phi_+}=W_{t\lambda}\otimes U_{t\lambda}\ket{\psi_{\rho}}.
  \end{align}
   where \(V_{t\lambda}\) denotes a unitary operator given by the polar decomposition of the state \(\sqrt{\rho}\sqrt{\rho _{t\lambda}}\) such that
  \begin{equation}
      \sqrt{\rho}\sqrt{\rho _{t\lambda}}=V_{t\lambda}\sqrt{\sqrt{\rho _{t\lambda}}\rho\sqrt{\rho _{t\lambda}}}.
  \end{equation}
  and \(W_{t\lambda}\) denotes a unitary operator defined as \(W_{t\lambda}=\left( U_{t\lambda}^{-1}V_{t\lambda}^{-1}. \right)^T\).
  Expanding $W_{t\lambda}$ for $t\lambda$, we express
  \begin{equation}
      W_{t\lambda}=I+\sum_{k=1}^N t\lambda_k A_k+\sum_{k,l=1}^N t^2\lambda_k\lambda_l B_{kl}+O(t^3)
  \end{equation}
  which implies
  \begin{equation}
      \begin{split}
      W_{t\lambda}^\dagger W_{t\lambda} = &I+t\sum_{k=1}^N \lambda_k(A_k+A_k^\dagger)\\
      &+t^2\left[ \sum_{k,l=1}^N \lambda_k \lambda_l (A_k^\dagger A_l+B_{kl}+B_{kl}^\dagger) \right]+O(t^3).
      \end{split}
  \end{equation}
  
  with $W_{t\lambda}$ unitary, we have
  \begin{align}
      &\sum_{k=1}^N \lambda_k(A_k+A_k^\dagger)=0\\
      &\sum_{k,l=1}^N \lambda_k \lambda_l \left(A_k^\dagger A_l+B_{kl}+B_{kl}^\dagger\right)=0.
  \end{align}
  From the first equation, with \(A_k\) a skew-Hermitian operator, we can write \(A_k=-i X_k^R\) using the Hermitian operator \(X_k^R\).
  From the second equation, by considering \(\lambda_k=\lambda_l=1\) and 0 otherwise, we obtain \(X_k^R X_l^R+X_l^R X_k^R+(B_{k l}+B_{l k})+(B_{k l}+B_{l k})^{\dagger}=0\).
  This implies
  \begin{equation}
      B_{k l}+B_{l k}=-\frac{X_k^R X_l^R+X_l^R X_k^R}{2}+2iC_{k,l}
  \end{equation}
  where $C_{k,l}$ is the Hermitian operator.
  Therefore, we describe
  \begin{equation}
      \begin{split}
      W_{t\lambda}=I-it\sum_{k=1}^N \lambda_k X_k^R &- \frac{t^2}{2} \left( \sum_{k=1}^N \lambda_k X_k^R \right)^2\\ 
      &+ it^2\sum_{k,l=1}^N C_{k,l}+O(t^3)
      \end{split}
  \end{equation}
  Expanding \(U_{t\lambda}\) for $t\lambda$, we describe
  \begin{equation}
      U_{t\lambda}=I-it\sum_{k=1}^N\lambda_k X_k^S-\frac{t^2}{2}\left(\sum_{k=1}^N\lambda_k X_k^S  \right)^2+O(t^3).
  \end{equation}
  Therefore, replacing \(X_S:=\sum_{k=1}^N\lambda_k X_k^S\), \(X_R:=\sum_{k=1}^N\lambda_k X_k^R\), and \(C=\sum_{k,l}\lambda_k\lambda_lC_{kl}\) for instances of a single conserved quantity, we have
  \begin{align}
      F(\rho,\rho_{\rho_{t\lambda}})=1-\frac{t^2}{2}\lambda^T\hat{V}_{\ket{\psi_\rho}} (\{X_n^S+X_n^R\}) \lambda +O(t^3).
  \end{align}
  Because the second-order derivative of fidelity is equal to the quantum Fisher information, we have
  \begin{align}
      \lambda ^T \hat{F}_\rho\left(\{X_n^S\}\right)\lambda &=F_\rho\left(\sum_{k=1}^N\lambda_k X_k^S\right)\nonumber\\
      &=8\lim_{t\to 0}\frac{1-F(\rho,\rho_{t\lambda})}{t^2}\nonumber\\
      &=4V_{\ket{\psi_\rho}}\left( \sum_{k=1}^N\lambda_k \left(X_k^S+X_k^R\right) \right)\nonumber\\
      &=4\lambda^T\hat{V}_{\ket{\psi_\rho}}\left(\{X_n^S+X_n^R\}\right)\lambda
  \end{align}
  which implies
  \begin{equation}
      \hat{F}_{\rho}\left(\{X_n^S\}\right)=4\times\hat{V}_{\ket{\psi_\rho}}\left(\{X_n^S+X_n^R\}\right).
  \end{equation}

\end{document}